\documentclass[aps,onecolumn,preprint,superscriptaddress,nofootinbib]{revtex4-1}

\usepackage[T1]{fontenc}
\usepackage[utf8]{inputenc}
\usepackage{mathtools}
\usepackage{graphicx}
\usepackage{caption}
\usepackage{subcaption}
\usepackage{amsmath,amssymb}
\usepackage[colorlinks=true, pdfstartview=FitV, linkcolor=blue, citecolor=blue, urlcolor=magenta, breaklinks=true]{hyperref}
\usepackage{orcidlink}
\usepackage{verbatim}

\begin{document}

\date{\today}

\title{The sound of quintessence: analogue Kiselev acoustic black holes }

\author{Luis C. N. Santos \orcidlink{0000-0002-6129-1820}}
\email{luis.santos@ufsc.br}

\affiliation{Departamento de F\'isica, CFM - Universidade Federal de Santa Catarina; \\ C.P. 476, CEP 88.040-900, Florian\'opolis, SC, Brazil.}

\author{H. S. Vieira \orcidlink{0000-0003-0909-2717}}
\email{horacio.santana.vieira@hotmail.com}
\email{horacio.santana-vieira@tat.uni-tuebingen.de}

\affiliation{Theoretical Astrophysics, Institute for Astronomy and Astrophysics, University of T\"{u}bingen, 72076 T\"{u}bingen, Germany}

\author{Franciele M. da Silva \orcidlink{0000-0003-2568-2901}} 
\email{franmdasilva@gmail.com}

\affiliation{Departamento de F\'isica, CFM - Universidade Federal de Santa Catarina; \\ C.P. 476, CEP 88.040-900, Florian\'opolis, SC, Brazil.}

\author{V. B. Bezerra \orcidlink{0000-0001-7893-0265}} 
\email{valdir@fisica.ufpb.br}

\affiliation{Departamento de F\'isica,- Universidade Federal da  Paraíba; \\ C.P. 5008, CEP 58.059-900, João Pessoa, PB, Brazil.}

\begin{abstract}
In this work, we demonstrate that the geometry of a spherically symmetric black hole surrounded by a Kiselev anisotropic fluid can be effectively mimicked by an experimental setup as the ones used to investigate some physical phenomena associated with acoustic black holes. Thus, we construct the metric describing Kiselev acoustic black holes by using the Gross--Pitaevskii theory and present a general analytical solution that encompasses, as particular cases, several classes of geometries associated with black holes. This unified framework allows for the description of a wide variety of analogue spacetimes, including new analogue geometries that have not been previously explored in the literature.  Then, we examine the behavior of scalar field perturbations in this background by solving the massless Klein--Gordon equation. Depending on the boundary conditions between the acoustic event horizon and infinity, we obtain the quasinormal and quasibound spectra. This study opens up avenues for experimental investigation within the context of analog gravity models, by offering new possibilities to simulate and study black hole phenomena in laboratory settings.
\end{abstract}

\maketitle

\section{Introduction}

The study of black holes has for a long time captivated physicists, providing a possibility to get some understanding of the aspects of quantum gravity and the nature of spacetime. However, direct experimental investigation of astrophysical black holes remains challenging due to their remote and extreme environments. On the other hand, analogue systems including acoustic black holes, which can be realized in the laboratory, replicate certain aspects of astrophysical black holes, and for this reason, emerged as a particularly promising avenue for exploring phenomena such as event horizons, black hole spectroscopy, superradiance, and Hawking radiation in laboratory settings \cite{Visser:1997ux,Fischer:2001jz,Kokkotas:1999bd,Nollert:1999ji,Konoplya:2011qq,Schutzhold:2002rf,Dolan:2011ti,Torres:2020tzs,Berti:2004ju,Cardoso:2004fi,Patrick:2018orp,Vieira:2019dau,Destounis:2023ruj,Rosato:2024arw,Pezzella:2024tkf,Jaramillo:2020tuu,Boyanov:2022ark,Boyanov:2023qqf,Cheung:2021bol,Courty:2023rxk,Boyanov:2024fgc,Patrick:2020baa,Torres:2016iee,Richartz:2014lda,Richartz:2012bd,Mascher:2022pku,Mollicone:2024lxy}.
Firstly proposed by Unruh in 1981 \cite{unruh1981}, the concept of an acoustic black hole involves the creation of a region within a fluid flow where the velocity exceeds the local speed of sound, effectively trapping phonons in a manner analogous to the trapping of light by an astrophysical black hole's event horizon. The key aspect associated with such analogue systems is the fact that the propagation of sound waves in such a medium can be described by equations formally equivalent to those governing scalar fields in a curved spacetime, thereby providing a fertile ground for theoretical and experimental investigations.  

A striking feature of analogue gravity is its generality: a wide variety of fluid systems \cite{Patrick:2019kis,Torres:2019sbr,Euve:2021mnj,Ciszak:2021xlw,Jannes:2010sa,Jacobson:1998ms} can exhibit black hole analogues, provided that the background flow can be cast in an effective Lorentzian geometry for perturbations. In classical hydrodynamics, the analogy arises naturally in barotropic, inviscid, and irrotational fluids, where linear perturbations satisfy a massless Klein--Gordon equation in a curved spacetime whose metric depends on the properties of the flow. Beyond classical fluids, the scope of analogue gravity has expanded dramatically with the incorporation of quantum fluids. In particular, Bose-Einstein condensates (BECs), governed at the mean-field level by the nonlinear Gross--Pitaevskii equation (GPE), offer an exceptionally rich platform for investigating acoustic black holes \cite{Garay:1999sk}. Furthermore, acoustic horizons have been theoretically and experimentally explored in a broad range of alternative media, including superfluid helium \cite{volovik2003universe}, nonlinear optical systems \cite{Philbin:2007ji,Belgiorno:2010wn}, and electric circuits \cite{Tokusumi:2018wii,Katayama:2021itj}. Each of these systems brings distinct physical characteristics that impact the form of the analogue metric, the nature of the excitations, and the feasibility of detecting analogue Hawking radiation.  

On the other hand, several astrophysical black hole geometries have been studied in the context of General Relativity (GR). Black holes characterized by an energy-momentum tensor linked to specific matter distributions include a notable class surrounded by a fluid known as the Kiselev black hole \cite{kiselev}, often referred to in the literature as a black hole with quintessence. This type of solution has been widely used to model various physical scenarios involving an anisotropic fluid equation of state \cite{santos2023kiselev,NunesdosSantos:2025alw,Santos:2024vby}. Depending on the choice for the equation of state parameter $\varpi$ are capable of describing the accelerated expansion observed in the universe. Moreover, Kiselev’s anisotropic fluid has been the subject of extensive research in different contexts, including black hole shadows \cite{shadow1,shadow2,shadow3,shadow4}, quasinormal modes \cite{quasi1,quasi2,quasi3,quasi4,quasi5,quasi6,quasi7,quasi8,quasi9}, black hole thermodynamics \cite{termo1,termo2,termo3,termo4,termo5,termo6,termo7,termo8,Lobo:2019jdz}, as well as in modified gravity frameworks \cite{heydarzade2017black,sakti_kerrnewmannutkiselev_2020,morais_thermodynamics_2022,gogoi2023joule,ghosh_rotating_2024,daSilva:2025cwz}. Additionally, the dynamics of Kiselev black holes have been explored in scenarios involving the collapse of null and timelike thin shells within this geometry \cite{saadati2021thin,javed2024impact,javed2024klein}.

One may naturally ask whether it is possible to reproduce, in an analogous acoustic medium, the behavior of black holes in GR surrounded by fluids. In this article, we address this question by proposing an acoustic analogue of the Kiselev black hole -- a solution that incorporates a surrounding fluid characterized by a general equation of state parameter. By constructing such an analogue model, we demonstrate that black hole configurations with various values of the parameter of the fluid's equation of state, including those corresponding to quintessence, can be effectively reproduced in a laboratory setting. In this way, it becomes possible to ``hear'' the sound of quintessence through an acoustic analogue setup.  Such analogue geometries may help to understand the effects of different types of matter contents on black hole structures and may motivate experimental designs aimed at testing fundamental aspects of analogue gravitation beyond the vacuum case.

This work is organized as follows. In Sec.~\ref{KABH}, we obtain the main equations describing an analogue of the Kiselev black hole in the context of a quantum fluid. In Sec.~\ref{SWE}, we explore some of the properties of the Kiselev acoustic black holes by analyzing the solutions of the massless Klein-Gordon equation. In Subsec.~\ref{QBSs}, we use the Vieira--Bezerra--Kokkotas (VBK) approach \cite{AnnPhys.373.28,Vieira:2021xqw} to obtain the quasibound states for three special cases: quintessence, dust matter and radiation. In Subsec.~\ref{QNMs}, by means of the Wentzel--Kramers--Brillouin (WKB) approximation \cite{Wentzel:1926aor,Kramers:1926njj,brillouin1926mecanique} we obtain the quasinormal modes associated with these same cases. Finally, in Sec.~\ref{conclusions} we summarize our main findings and conclusions.

\section{The Kiselev acoustic black holes}\label{KABH}

We start this study by considering the action of the Gross--Pitaevskii (GP) theory that describes a nonlinear complex scalar field, $\varphi$, which reads \cite{NuovoCimento.20.454,JETP.13.451}
\begin{equation}
\mathcal{S}_{\rm GP}=\int d^{4}x\ \sqrt{-g}\ \biggl(|\partial_{\mu}\varphi|^{2}+m^{2}|\varphi|^{2}-\frac{b}{2}|\varphi|^{4}\biggr),
\label{eq:action_GP}
\end{equation}
where $b$ and $m$ are a coupling constant and a temperature-dependent parameter, respectively. In fact, $m$ depends on the Hawking--Unruh temperature $T$ of the resultant acoustic solution and hence it encodes both the information of the black holes and the acoustic metric. Thus, for our purposes, we assume this temperature dependence as $m^{2} \sim T-T_{c}$, where $T_{c}$ is the critical temperature of the theory that describes phase transitions, with $\varphi$ being the corresponding order parameter. Therefore, when $T>T_c$ the phenomenological parameter $m^2$ is positive, while for $T<T_c$ it becomes negative. For $T=T_c$, we have $m^2=0$. Then, from this action, we can derive the equation of motion with respect to $\varphi$, which is given by
\begin{equation}
\Box\varphi+m^{2}\varphi-b|\varphi|^{2}\varphi=0.
\label{eq:equation_motion_GP}
\end{equation}
It is well known that one can obtain an acoustic black hole solution within the GP theory by considering perturbations of the complex scalar field around a fixed spacetime background. To do this, we choose the spacetime background metric as
\begin{equation}
ds^{2}=g_{tt}\ dt^{2}+g_{rr}\ dr^{2}+g_{\vartheta\vartheta}\ d\vartheta^{2}+g_{\phi\phi}\ d\phi^{2}.
\label{eq:background_metric}
\end{equation}
Next, we use the Madelung representation to write the complex scalar field as
\begin{equation}
\varphi=\sqrt{\rho(\mathbf{x},t)}\mbox{e}^{i\theta(\mathbf{x},t)},
\label{eq:Madelung_representation}
\end{equation}
where $\rho$ and $\theta$ are the fluid density and the phase respectively. They are related to the background solution in the fixed spacetime (denoted by subscript 0), as well as to the fluctuations (denoted by subscript 1), as
\begin{eqnarray}
\rho & = & \rho_{0}+\epsilon\rho_{1},\\
\theta & = & \theta_{0}+\epsilon\theta_{1}.
\label{eq:fluid_phase_fluctuations}
\end{eqnarray}
Thus, from the equation of motion (\ref{eq:equation_motion_GP}), we obtain two equations by working in the long-wavelength limit and neglecting the quantum potential terms; the leading order equation for the background fluid density reads
\begin{equation}
b\rho_{0}=m^{2}-g^{\mu\nu}v_{\mu}v_{\nu},
\label{eq:leading_order}
\end{equation}
and the sub-leading order equation for the phase fluctuations reads
\begin{equation}
\frac{1}{\sqrt{\mathcal{-G}}}\partial_{\mu}(\sqrt{\mathcal{-G}}\mathcal{G}^{\mu\nu}\partial_{\nu}\theta_{1})=0.
\label{eq:phase_fluctuations}
\end{equation}
Here, we defined the background fluid four-velocity as $v_{\mu}=(-\partial_{t}\theta_{0},\partial_{i}\theta_{0})$, with $i=(r,\vartheta,\phi)$. Note that we obtained a relativistic wave equation that is similar to the massless Klein-Gordon equation, which governs the propagation of the phase fluctuations as weak excitations in a homogeneous stationary condensate, where $\mathcal{G}=\det(\mathcal{G}_{\mu\nu})$ and $\theta_{1}=\theta_{1}(t,r,\vartheta,\phi)$. The effective metric $\mathcal{G}_{\mu\nu}$ is extracted from the massless Klein-Gordon equation (\ref{eq:phase_fluctuations}), and it can be written as
\begin{equation}
\mathcal{G}_{\mu\nu}=\frac{c_s}{\sqrt{c^2_s-v_{\mu}v^{\mu}}}
\begin{pmatrix}
g_{tt}(c^2_s- v_i v^i) & \vdots & -v_{t}v_{i}\cr
\cdots\cdots\cdots\cdots & \cdot &\cdots\cdots\cdots\cdots\cdots\cdots\cr
 -v_{i}v_{t} & \vdots & g_{ii}(c^2_s-v_\mu v^\mu)\delta^{ij}+v_i v_j\cr
\end{pmatrix}\!,
\label{eq:effective_metric}
\end{equation}
where we have defined the speed of sound as $c_{s}^{2} \equiv b\rho_{0}/2$. Finally, under the assumptions that $v_{t} \neq 0$, $v_{r} \neq 0$, $v_{\vartheta}=v_{\phi} = 0$ without any time dependence and $g_{rr}g_{tt}=-1$, together with the coordinate transformation
\begin{equation}
dt \rightarrow dt+\frac{v_{r}v_{t}}{g_{tt}(c_{s}^{2}-v_{i}v^{i})}dr,
\label{eq:coordinate_transformations}
\end{equation}
we can write the line element for static acoustic black holes as
\begin{equation}
ds^{2}=c_{s}\sqrt{c_{s}^{2}-v_{\mu}v^{\mu}}\biggl(\frac{c_{s}^{2}-v_{r}v^{r}}{c_{s}^{2}-v_{\mu}v^{\mu}}g_{tt}\ dt^{2}+\frac{c_{s}^{2}}{c_{s}^{2}-v_{r}v^{r}}g_{rr}\ dr^{2}+g_{\vartheta\vartheta}\ d\vartheta^{2}+g_{\phi\phi}\ d\phi^{2}\biggr).
\label{eq:acoustic_metric}
\end{equation}
At this point, by choosing both the spacetime background and the fluid four-velocity components, we can completely characterize the acoustic spacetime given by Eq.~(\ref{eq:acoustic_metric}). This will be done in what follows.

We are going to focus on the Minkowski flat spacetime with a radial velocity intended to mimic the Kiselev black hole background. Thus, the spacetime background metric in spherical coordinates is given by \cite{MTW:1973}
\begin{eqnarray}
ds^{2} = dt^{2}+dr^{2}+r^{2}\ d\vartheta^{2}+r^{2}\sin^{2}\vartheta\ d\phi^{2},
\label{eq:Minkowski_metric}
\end{eqnarray}
while the radial component of the fluid four-velocity can be written as
\begin{equation}
v_{r} \sim \sqrt{\frac{D}{r}-\frac{C_{\varpi}}{r^{3\varpi+1}}},
\label{eq:radial_fluid_component}
\end{equation}
where $D$ and $C_{\varpi}$ can be identified, without loss of generality, as the draining
and charge parameters, respectively. This radial velocity can be understood as follows. When the parameter $C_{\varpi}$ disappears, this radial component could describe the escape velocity of an observer who maintains a stationary position at the radial position $r$; from an experimental point of view, it may describe a radial (draining) fluid flow. On the other hand, for different values of the anisotropic fluid parameter $\varpi$, we can obtain different interpretations for the charge parameter; here, we will consider $\varpi=(-2/3,0,1/3)$, which means a ``charge'' due to quintessence, dust matter, and radiation, respectively. In which concerns to the four-velocity, we first rescale $m^{2} \rightarrow m^{2}/2c_{s}^{2}$ and $v_{\mu}v^{\mu} \rightarrow v_{\mu}v^{\mu}/2c_{s}^{2}$, and then by working in the limit of critical temperature of the GP theory, which implies that $m^{2}\rightarrow0$, we get $v_{\mu}v^{\mu}=-1$. Thus, in order to fulfill the relation $v_{\mu}v^{\mu}=-1$, the temporal component of the fluid four-velocity should be
\begin{equation}
v_{t}=\sqrt{1+\frac{D}{r}-\frac{C_{\varpi}}{r^{3\varpi+1}}}.
\label{eq:temporal_fluid_component}
\end{equation}

Therefore, we can rewrite the line element given by Eq.~(\ref{eq:acoustic_metric}), as
\begin{equation}
ds^{2}=\sqrt{3}c_{s}^{2}\biggl[-f(r)\ dt^{2}+\frac{1}{f(r)}\ dr^{2}+r^{2}\ d\vartheta^{2}+r^{2}\sin^{2}\vartheta\ d\phi^{2}\biggr],
\label{eq:LTABH_metric}
\end{equation}
where the acoustic metric function $f(r)$ is given by
\begin{equation}
f(r)=1-\frac{D}{r}+\frac{C_{\varpi}}{r^{3\varpi+1}}.
\label{eq:f(r)_KABH}
\end{equation}
Thus, this line element describes Kiselev acoustic black holes (KABHs). For simplicity and without loss of generality, we will fix $c_{s}^{2}=1/\sqrt{3}$. It is worth noticing that the acoustic event horizons depend on the choice of the anisotropic fluid parameter $\varpi$, as they are solutions of the surface equation given by
\begin{equation}
f(r)=0.
\label{eq:surface_equation}
\end{equation}
Since the outer acoustic event horizon is the last surface from which sound waves could still escape the acoustic black hole, it is meaningful to study the motion of massless scalar particles propagating at the external region of the KABH spacetime, which will be presented in the next section.

\section{Scalar wave equation}\label{SWE}

We are interested in some basic characteristics of these Kiselev acoustic black holes, in particular the ones related to the classical scalar wave scattering such as quasinormal modes (QNMs) and quasibound states (QBSs). To perform this analysis, we must consider the minimally coupled massless scalar field as a probe, whose covariant equation of motion is given by Eq.~(\ref{eq:phase_fluctuations}). Thus, to obtain analytical solutions of the massless Klein-Gordon equation (\ref{eq:phase_fluctuations}), and due to stationarity and axisymmetry, we use the following separation ansatz
\begin{equation}
\theta_{1}(t,r,\vartheta,\phi) \simeq \mbox{e}^{-i \omega t}U(r)P(\vartheta)\mbox{e}^{i m \phi},
\label{eq:ansatz}
\end{equation}
where $U(r)=R(r)/r^N$ is the radial function with $N \in \mathbb{Z}$, $P(\vartheta)$ is the polar angle function, $m$ $(\in \mathbb{Z})$ is the magnetic or azimuthal quantum number, and $\omega$ is the frequency (energy, in the natural units).

Then, by substituting Eq.~(\ref{eq:ansatz}) into Eq.~(\ref{eq:phase_fluctuations}), we obtain two ordinary differential equations, namely,
\begin{equation}
P^{\prime\prime}(\vartheta)+\frac{\cos\vartheta}{\sin\vartheta}P^{\prime}(\vartheta)+\biggl(\lambda-\frac{m^{2}}{\sin^{2}\vartheta}\biggr)P(\vartheta)=0,
\label{eq:angular_equation}
\end{equation}
and
\begin{eqnarray}
&& R^{\prime\prime}(r)+\biggl[\frac{2(1-N)}{r}+\frac{f^\prime(r)}{f(r)}\biggr]R^{\prime}(r)\nonumber\\
&& +\frac{1}{f^2(r)}\biggl\{\omega^2+\frac{f(r)}{r^2}[N^2f(r)-Nf(r)-\lambda-Nrf^\prime(r)]\biggr\}R(r)=0,
\label{eq:radial_equation}
\end{eqnarray}
where $\lambda$ is the separation constant, and prime denotes differentiation of the polar and radial functions with respect to $\vartheta$ and $r$, respectively. The analytical solution of the polar equation (\ref{eq:angular_equation}) is given in terms of the associated Legendre functions $P(\vartheta)=P_{\nu}^{m}(\cos\vartheta)$ with degree $\nu$ $(\in \mathbb{C})$ and order $m \geq 0$ $\in \mathbb{Z}$, such that $\lambda=\nu(\nu+1)$. On the other hand, in the radial equation, which remains to be solved according to the value of $\varpi$, the choice of $N$ is made only for convenience. In order to study the QNMs is more convenient to set $N=1$ since it allows us to write Eq.~(\ref{eq:radial_equation}) in a Schr\"odinger-like form, while to study the QBSs is more convenient to set $N=0$ since it allows us to express Eq.~(\ref{eq:radial_equation}) in a Heun-type form \cite{Ronveaux:1995}.

\subsection{Quasibound states}\label{QBSs}

The VBK approach is a powerful analytical method designed to study quasibound states of black holes, particularly those involving massive fields such as scalar, vector, or even tensor perturbations. This method utilizes the mathematical framework of Heun functions to solve the wave equations governing field dynamics in black hole spacetimes. Quasibound states are characterized by complex frequencies, representing oscillatory modes that are trapped near the black hole and decay slowly over time as a result of energy leakage through the potential barrier. The VBK method excels at identifying these modes by accurately analyzing the boundary conditions of the problem -- ingoing waves at the (exterior) black hole event horizon and exponential decay at spatial infinity. This approach is particularly useful in scenarios where standard numerical methods become difficult to apply, offering a clear path to extracting physical insights in spacetimes such as Schwarzschild, Kerr, and Kerr--Newman geometries. Furthermore, the method plays an important role in understanding phenomena like superradiant instabilities, which occur when rotating black holes interact with massive scalar fields. Overall, the VBK formalism stands out for its analytical depth and adaptability, providing a crucial tool for probing the rich structure of black hole perturbation spectra.

In summary, quasibound states are resonant frequencies living at the potential barrier wall. They are eigensolutions of a boundary condition problem. Therefore, QBSs describe bounded particles around a black hole spacetime background. These eigenfrequencies constitute one of the components need to a spectral study of field perturbations in astrophysical and analog black holes.

In this sense, to obtain the QBS spectrum, we can set $N=0$ and then rewrite Eq.~(\ref{eq:radial_equation}) as
\begin{equation}
R^{\prime\prime}(r)+\biggl[\frac{2}{r}+\frac{f^\prime(r)}{f(r)}\biggr]R^{\prime}(r)+\biggl[\frac{\omega^2}{f^2(r)}-\frac{\lambda}{r^2f(r)}\biggr]R(r)=0.
\label{eq:radial_equation_QBSs}
\end{equation}
Now, we are going to transform Eq.~(\ref{eq:radial_equation_QBSs}) in a Heun-type equation. We will do this for each value of the parameter $\varpi$.

\subsubsection{Quintessence}\label{q}

\begin{figure}[ht]
\centering
\includegraphics[scale=1]{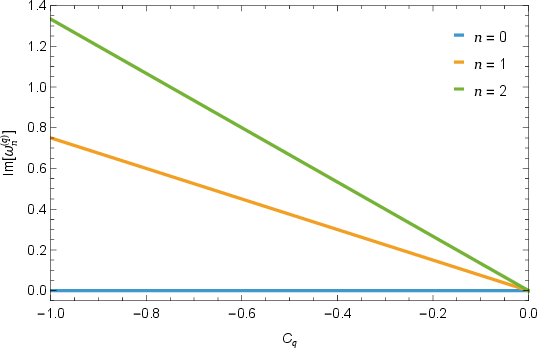}
\caption{QBSs for the quintessence case as a function of the charge parameter $C_{q}$.}
\label{fig:Fig1_QBSs_KABH}
\end{figure}

For the quintessence component, the anisotropic fluid parameter is $\varpi_{q}=-2/3$. In this case, the metric function $f(r)$ reads
\begin{equation}
f(r)=1-\frac{D}{r}+C_{q}r,
\label{eq:f(r)_q}
\end{equation}
such that the surface equation has two roots, namely,
\begin{equation}
r_{+}=\frac{-1+\sqrt{1+4C_{q}D}}{2C_{q}},
\label{eq:ext_q}
\end{equation}
and
\begin{equation}
r_{-}=\frac{-1-\sqrt{1+4C_{q}D}}{2C_{q}},
\label{eq:int_q}
\end{equation}
which can be identified, without loss of generality, as the exterior and the interior acoustic event horizons, respectively. By following the VBK approach (we cordially invite the reader to see Ref.~\cite{Vieira:2023ylz}), we obtain an analytical solution for the radial Klein--Gordon equation (\ref{eq:radial_equation_QBSs}) given in terms of the general Heun functions, namely,
\begin{equation}
R(x)=C_{1}\ x^{\frac{1}{2}(\gamma-1)}(x-x_{1})^{\frac{1}{2}(\epsilon-1)}\mbox{HeunG}(x_{1},\xi,\alpha,\beta,\gamma,\delta,x),
\label{eq:radial_solution_q}
\end{equation}
where $C_{1}$ is a constant, and the new radial coordinate $x$ is defined as
\begin{equation}
x=\frac{r-r_{-}}{-r_{-}},
\label{eq:x_solution_q}
\end{equation}
with the parameters $x_{1}$, $\xi$, $\alpha$, $\beta$, $\gamma$, $\delta$ and $\epsilon$ being given by
\begin{eqnarray}
x_{1} & = & \frac{r_{+}-r_{-}}{-r_{-}}\label{eq:x1_q},\\
\xi & = & -\frac{\lambda+2i\omega r_{-}}{C_qr_{-}}\label{eq:xi_q},\\
\alpha & = & 1-\frac{i\omega-\sqrt{C_q^{2}-\omega^{2}}}{C_q}\label{eq:alpha_q},\\
\beta & = & 1-\frac{i\omega+\sqrt{C_q^{2}-\omega^{2}}}{C_q}\label{eq:beta_q},\\
\gamma & = & 1+\frac{2i\omega r_{-}}{C_q(r_{+}-r_{-})}\label{eq:gamma_q},\\
\delta & = & 1\label{eq:delta_q},\\
\epsilon & = & 1-\frac{2i\omega r_{+}}{C_q(r_{+}-r_{-})}\label{eq:epsilon_q}.
\end{eqnarray}
Then, from the polynomial $\alpha$-condition we obtain the following expression for the QBS frequencies
\begin{equation}
\omega^{(q)}_{n}=-i\frac{C_{q}n(n+2)}{2(n+1)},
\label{eq:QBSs_q}
\end{equation}
where $n(=0,1,2,\ldots)$ is the overtone number. In Fig.~\ref{fig:Fig1_QBSs_KABH} we present the behavior of these QBSs as a function of the charge parameter $C_{q}$, for different overtone numbers $n$.

\subsubsection{Dust Matter}\label{m}

\begin{figure}[ht]
\centering
\includegraphics[scale=1]{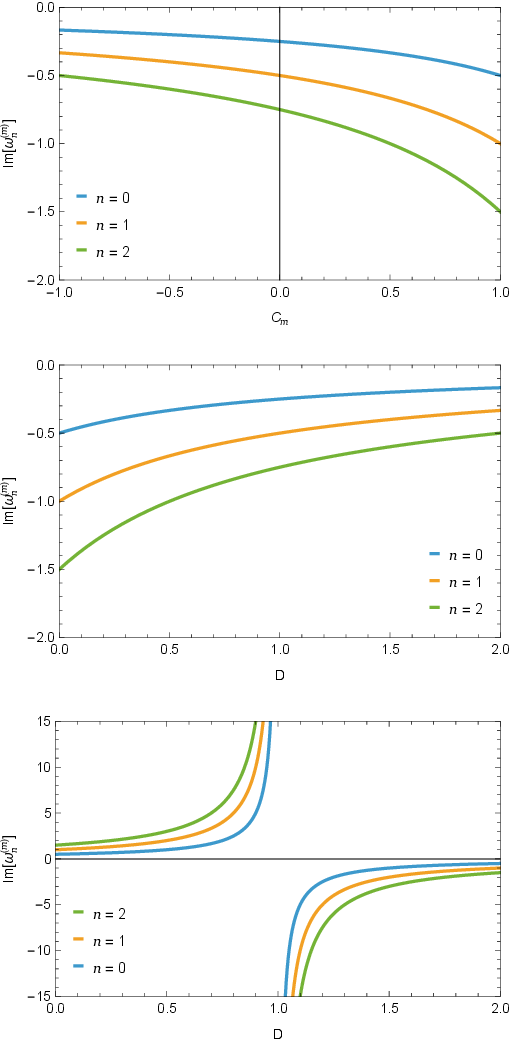}
\caption{QBSs for the dust matter case. Top panel: As a function of the charge parameter $C_{m}$, with $D=2$. Middle panel: As a function of the draining parameter $D$, with $C_{m}=-1$. Bottom panel: As a function of the draining parameter $D$, with $C_{m}=+1$.}
\label{fig:Fig2_QBSs_KABH}
\end{figure}

Next, for the dust matter component, the anisotropic fluid parameter is $\varpi_{m}=0$. In this case, the metric function $f(r)$ reads
\begin{equation}
f(r)=1-\frac{D-C_{m}}{r},
\label{eq:f(r)_m}
\end{equation}
such that the surface equation has one root, namely,
\begin{equation}
r_{h}=D-C_{m},
\label{eq:ext_m}
\end{equation}
which can be identified, without loss of generality, as the radius of the acoustic event horizon. Again, by following the VBK approach (we cordially invite the reader to see the Appendix A of Ref.~\cite{PhysRevD.111.104025}), we obtain an analytical solution for the radial Klein--Gordon equation (\ref{eq:radial_equation_QBSs}) given in terms of the confluent Heun functions, namely,
\begin{equation}
R(x)=C_{1}\ (x-1)^{\frac{\delta}{2}}\mbox{e}^{\frac{\alpha}{2}}\mbox{HeunC}(\alpha,\beta,\gamma,\delta,\eta;x),
\label{eq:radial_solution_m}
\end{equation}
where $C_{1}$ is a constant, and the new radial coordinate $x$ is defined as
\begin{equation}
x=\frac{r}{r_{h}},
\label{eq:x_solution_m}
\end{equation}
with the parameters $\alpha$, $\beta$, $\gamma$, $\delta$ and $\eta$ being given by
\begin{eqnarray}
\alpha & = & 2ir_{h}\omega\label{eq:alpha_m},\\
\beta & = & 0\label{eq:beta_m},\\
\gamma & = & -2ir_{h}\omega\label{eq:gamma_m},\\
\delta & = & 2r_{h}^{2}\omega^{2}\label{eq:delta_m},\\
\eta & = & -\lambda\label{eq:eta_m}.
\end{eqnarray}
Then, from the polynomial $\delta$-condition we obtain the following expression for the QBS frequencies
\begin{equation}
\omega^{(m)}_{n}=i\frac{n+1}{2(D-C_{m})},
\label{eq:QBSs_m}
\end{equation}
where $n(=0,1,2,\ldots)$ is the overtone number. In Fig.~\ref{fig:Fig2_QBSs_KABH} we present the behavior of these QBSs as a function of the charge $C_{m}$ and draining $D$ parameters, for different overtone numbers $n$.

\subsubsection{Radiation}\label{r}

\begin{figure}[ht]
	\centering
	\includegraphics[scale=1]{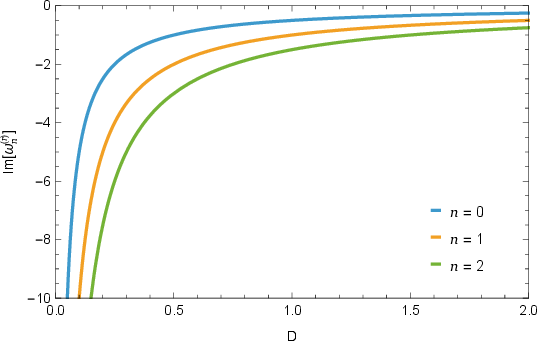}
	\caption{QBSs for the radiation case as a function of the draining parameter $D$.}
	\label{fig:Fig3_QBSs_KABH}
\end{figure}

Finally, for the radiation component, the anisotropic fluid parameter is $\varpi_{r}=1/3$. In this case, the metric function $f(r)$ reads
\begin{equation}
f(r)=1-\frac{D}{r}+\frac{C_{r}}{r^2},
\label{eq:f(r)_r}
\end{equation}
such that the surface equation has two roots, namely,
\begin{equation}
r_{+}=\frac{D+\sqrt{D^2-4C_{r}}}{2},
\label{eq:ext_r}
\end{equation}
and
\begin{equation}
r_{-}=\frac{D-\sqrt{D^2-4C_{r}}}{2},
\label{eq:int_r}
\end{equation}
which can be identified, without loss of generality, as the exterior and the interior acoustic event horizons, respectively. Once more, by following the VBK approach, we obtain an analytical solution for the radial Klein--Gordon equation (\ref{eq:radial_equation_QBSs}) given in terms of the confluent Heun functions, namely,
\begin{equation}
R(x)=C_{1}\ x^{\frac{\beta}{2}}(x-1)^{\frac{\gamma}{2}}\mbox{e}^{\frac{\alpha}{2}}\mbox{HeunC}(\alpha,\beta,\gamma,\delta,\eta;x),
\label{eq:radial_solution_r}
\end{equation}
where $C_{1}$ is a constant, and the new radial coordinate $x$ is defined as
\begin{equation}
x=\frac{r-r_{-}}{r_{+}-r_{-}},
\label{eq:x_solution_r}
\end{equation}
with the parameters $\alpha$, $\beta$, $\gamma$, $\delta$ and $\eta$ being given by
\begin{eqnarray}
\alpha & = & 2i(r_{+}-r_{-})\omega\label{eq:alpha_r},\\
\beta & = & \frac{2ir_{-}^{2}\omega }{r_{+}-r_{-}}\label{eq:beta_r},\\
\gamma & = & -\frac{2ir_{+}^{2}\omega }{r_{+}-r_{-}}\label{eq:gamma_r},\\
\delta & = & 2(r_{+}^{2}-r_{-}^{2})\omega\label{eq:delta_r},\\
\eta & = & \frac{2 r_{-}^3 \omega ^2 (r_{-}-2 r_{+})}{(r_{+}-r_{-})^2}-\lambda\label{eq:eta_r}.
\end{eqnarray}
Then, from the polynomial $\delta$-condition we obtain the following expression for the QBS frequencies
\begin{equation}
\omega^{(r)}_{n}=-i\frac{n+1}{2D},
\label{eq:QBSs_r}
\end{equation}
where $n(=0,1,2,\ldots)$ is the overtone number. In Fig.~\ref{fig:Fig3_QBSs_KABH} we present the behavior of these QBSs as a function of the draining parameter $D$, for different overtone numbers $n$.

\subsection{Quasinormal modes}\label{QNMs}
QNMs play an important role in understanding the dynamical response of black holes to external perturbations. These modes represent characteristic oscillations with complex frequencies, whose real part determines the oscillation rate and the imaginary part governs the damping due to energy loss. In analogue gravitational systems, such as acoustic black holes, effective geometries emerging in fluid dynamics simulate key features of gravitational black holes. This subsection focuses on the scalar QNMs in acoustic black hole backgrounds, employing the WKB approximation to compute the associated frequencies using an automatic code \cite{Konoplya:2019hlu} to find QNMs. We start observing that for $N = 1$, Eq.~(\ref{eq:radial_equation}) can be written as
\begin{equation}
    \frac{d^2 R(r)}{dr_*^2} + \left[ \omega^2 - V(r) \right] R(r) = 0,
\end{equation}
where the (effective) potential is given by
\begin{equation}
    V(r) = f(r) \left[ \frac{\nu(\nu+1)}{r^2} + \frac{f'(r)}{r} \right] =\frac{\left( C + r^{3 \varpi} \left(r -D\right) \right) \left(r^{3 \varpi} \left( D + r \nu (1 + \nu) \right) -C(1 + 3 \varpi)\right)}{ r^{4 + 6 \varpi} }
,
\label{efetivo}
\end{equation}
with $r_*$ being the tortoise coordinate related to the metric function in the form $dr/dr_* = f(r)$. In this way, $r_* \rightarrow \infty$ is associated with $r \rightarrow \infty$ or de Sitter horizon and $r_* \rightarrow -\infty$ corresponds to the event horizon of the black hole geometry. To determine the QNMs, we employ the sixth-order WKB approximation \cite{Konoplya:2003ii}, in which the formula for the frequencies is related to the effective potential (Eq.~\ref{efetivo}) as follows:
\begin{equation}
    \frac{w^2 - V_0}{\sqrt{-2V_0^{''}}} -\Lambda_2 - \Lambda_3 -\Lambda_4 -\Lambda_5 -\Lambda_6 = n + \frac{1}{2},
\end{equation}
where $V_0$ is the point of maximum of the potential and $V_0^{''}$ the second derivative of the potential at this point. The explicit formula for the functions $\Lambda_i, \:\:  2 \leq i \leq 6$ is given in \cite{Iyer:1986np,Konoplya:2003ii} and an automatic code which finds QNMs can be found in \cite{Konoplya:2019hlu}.

\subsubsection{Quintessence}

\begin{figure}[ht]
	\centering
	\includegraphics[scale=0.95]{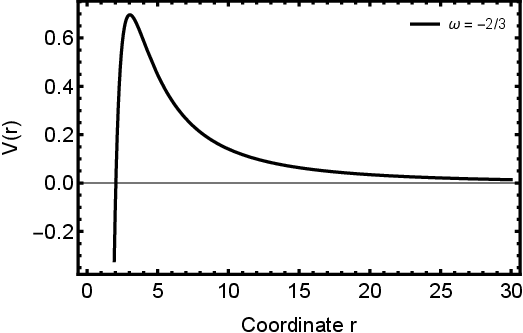}
	\caption{Effective potential for the quintessence case as a function of the radial coordinate, considering $n=4,C_q=-0.01,D=2,\nu=4$ and $\varpi = -2/3$.}
	\label{fig:FIG1_MQNs}
\end{figure}

\begin{figure}[ht]
	\centering
	\includegraphics[scale=0.9]{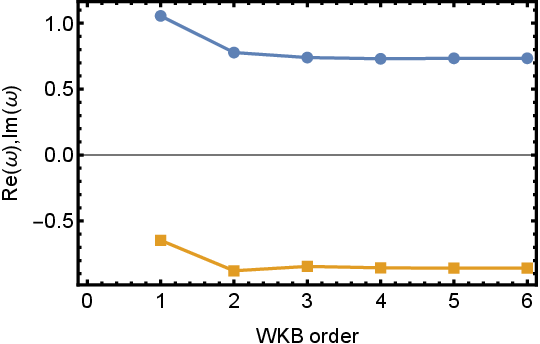}
	\caption{Here we show the convergence of the real (blue) and imaginary (yellow) parts of the QNMs as the order of the WKB approximation increases for the quintessence case. In this figure we assume $n=4,C_q=-0.01,D=2,\nu=4$ and $\varpi = -2/3$.}
	\label{fig:FIG2_MQNs}
\end{figure}

\begin{figure}[ht]
	\centering
	\includegraphics[scale=0.95]{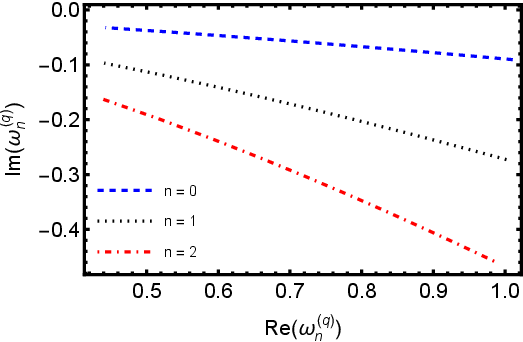}
	\caption{Behavior of the imaginary and real parts of the QNMs for the quintessence case, for varying charge parameter $C_q = -0.01,...,-0.1$, with step $0.01$ from left to right. We consider three different values of the overtone number and $D=2,\nu=5,\varpi = -2/3$.}
	\label{fig:FIG3_MQNs}
\end{figure}

We now turn our attention to the study the particular case of quintessence, corresponding to $\varpi = -2/3$. As discussed previously, this solution features both an inner and an outer horizon. The effective potential is plotted in Fig.~\ref{fig:FIG1_MQNs}, where we can clearly identify the region where it reaches its maximum, around $r \approx 3$.
We applied the sixth-order WKB method to the effective potential associated with this geometry. As shown in Fig.~\ref{fig:FIG2_MQNs}, from the second order onward, the real and imaginary parts begin to converge. The negative value of the imaginary part indicates the damping of the perturbation.
Note that $C_q$ must be a negative number in this solution in order to ensure a positive energy density in the astrophysical system. Accordingly, we preserve this condition for the analog system and compute the QNMs for several values of $C_q$. The results are shown in Fig.~\ref{fig:FIG3_MQNs}.
We fixed the draining parameter and varied the charge according to $C_q = -0.01, \ldots, -0.1$, with a step size of $0.01$. As can be seen, the imaginary part tends toward more negative values as both $C_q$ and $n$ increase, while the real part increases in magnitude. More negative values indicate faster decay of the perturbation, meaning the system returns to equilibrium more rapidly. The observed trend, in which the imaginary part becomes increasingly negative with larger values of $|C_q|$ and overtone number $n$ suggests that the presence of quintessence enhances the damping of the field. The real part of the quasinormal frequencies that increase with $|C_q|$ implies that the presence of quintessence leads to higher-frequency oscillations. 

\subsubsection{Dust Matter}

\begin{figure}[ht]
	\centering
	\includegraphics[scale=0.95]{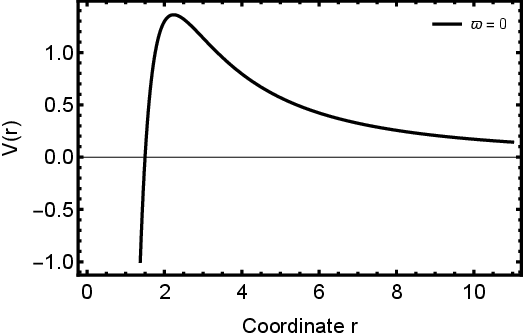}
	\caption{Effective potential for the dust matter case as a function of the radial coordinate, considering $n=4,C_m=0.5,D=2,\nu=4$ and $\varpi = 0$.}
	\label{fig:FIG4_MQNs}
\end{figure}

\begin{figure}[ht]
	\centering
	\includegraphics[scale=0.9]{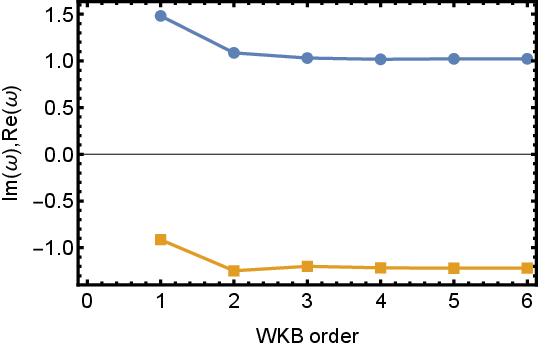}
	\caption{Here we show the convergence of the real (blue) and imaginary (yellow) parts of the QNMs as the order of the WKB approximation increases for the dust matter case. In this figure we assume $n=4,C_m=0.5,D=2,\nu=4$ and $\varpi = 0$.}
	\label{fig:FIG5_MQNs}
\end{figure}

\begin{figure}[ht]
	\centering
	\includegraphics[scale=0.9]{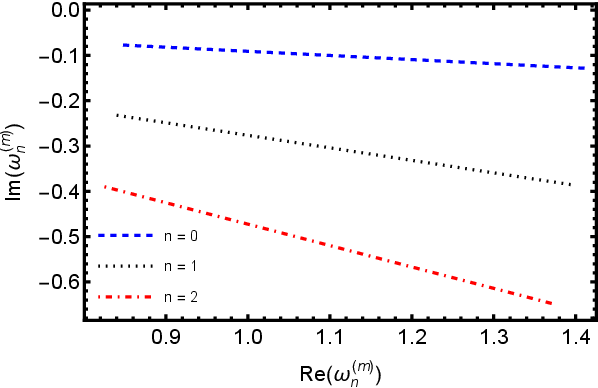}
	\caption{Behavior of the imaginary and real parts of the QNMs for the dust matter case, for varying charge parameter $C_m = -0.5,...,0.5$, with step $0.1$ from left to right. We consider three different values of the overtone number and $D=2,\nu=5,\varpi = 0$.}
	\label{fig:FIG6_MQNs}
\end{figure}

For astrophysical black holes, the value $\varpi = 0$ corresponds to a solution associated with dust matter. For the solution obtained here, which describes analog black holes, we analyze the behavior of the analog geometry for this specific case. In particular, we investigate QNMs associated with the analog dust matter black hole considering values for the draining parameter and the charge. The preliminary analysis indicates the existence of a single horizon for this geometry at $r_h =D - C_m$. Then, for the set of parameters used in Fig.~\ref{fig:FIG4_MQNs}, the horizon is placed at $r_h = 1.5$. The effective potential is shown in Fig.~\ref{fig:FIG4_MQNs}, where its maximum, for a particular set of parameter values, occurs around $r \approx 2.2$.     
Of course, the exact point of the maximum can be determined by evaluating the first derivative of the potential. We used this value to compute QNMs for the dust matter case. Figure \ref{fig:FIG5_MQNs} presents the convergence analysis with respect to the order of the WKB method, where it can be seen that from the third order onward, the results exhibit good convergence behavior.
In Fig.~\ref{fig:FIG6_MQNs}, we show the quasinormal modes for a fixed value of the draining parameter and increasing values of $C_m$. For the set of parameter values considered, it is evident that both the real and imaginary parts exhibit behavior similar to the quintessence case. That is, the imaginary part becomes more negative, while the real part increases as $C_m$ grows. 

\subsubsection{Radiation}
\begin{figure}[h]
	\centering
	\includegraphics[scale=0.95]{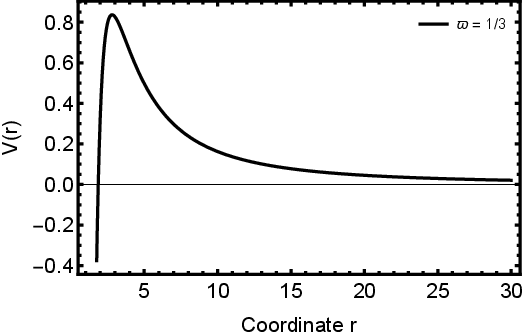}
	\caption{Effective potential for the radiation case as a function of the radial coordinate, considering $n=0,C_r=0.25,D=2,\nu=4$ and $\varpi = 1/3$.}
	\label{fig:FIG7_MQNs}
\end{figure}
\begin{figure}[h]
	\centering
	\includegraphics[scale=0.9]{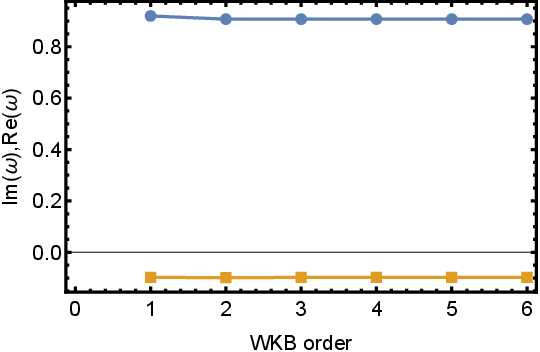}
	\caption{Here we show of the real (blue) and imaginary (yellow) parts of the QNMs as the order of the WKB approximation increases for the radiation case. In this figure we assume $n=0,C_r=0.25,D=2,\nu=4$ and $\varpi = 1/3$.}
	\label{fig:FIG8_MQNs}
\end{figure}
\begin{figure}[h]
	\centering
	\includegraphics[scale=0.9]{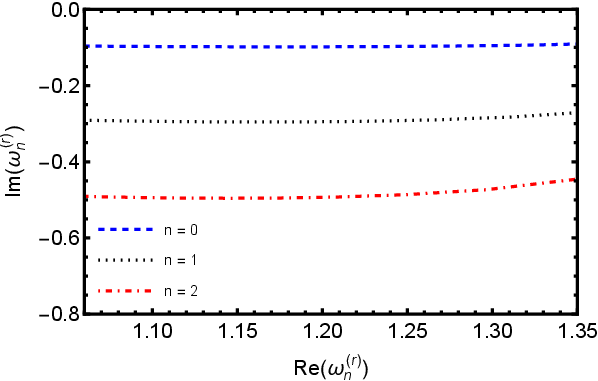}
	\caption{Behavior of the imaginary and real parts of the QNMs for the radiation case, for varying charge parameter $C_r = 0,...,1$, with step $0.1$ from left to right. We consider three different values of the overtone number and $D=2,\nu=5,\varpi = 1/3$.}
	\label{fig:FIG9_MQNs}
\end{figure}

Assuming that $\varpi = 1/3$, we recover the particular solution associated with radiation. This solution effectively describes a metric associated with a charged black hole with charge $Q$ in the astrophysical case. In our solution, we interpret $C_r$ as a factor that modifies the radial velocity component of the fluid. Depending on the value of $C_r$, the fluid velocity may either increase or decrease. In this particular case, the presence of a term $\sim 1/r^2$ provides a significant contribution for small values of $r$. In the astrophysical case, the charge appears squared in the geometry, so the term $\sim 1/r^2$ is always positive. In the acoustic case, however, the value of $C_r$ can, in principle, take negative values. Nevertheless, we will avoid such values in order to ensure a positive energy density. This metric features an outer horizon $r_+$ and an inner horizon $r_-$. In fact, $r_-$ is a Cauchy horizon. If $C_r = 0$, we recover the Schwarzschild solution with a single event horizon, and for $C_r > D^2$ the solution has no real horizons, leading to a naked singularity.

The effective potential is shown in Fig.~\ref{fig:FIG7_MQNs} and exhibits a well-defined behavior around its maximum point, similar to the previously discussed particular cases. Regarding the convergence of the quasinormal modes, Fig.~\ref{fig:FIG8_MQNs} shows convergence starting from the first order of the WKB method. In Fig.~\ref{fig:FIG9_MQNs}, the QNMs are displayed for varying charge parameter $C_r = 0, \ldots, 1,$ with a step of 0.1 and for $D = 2$. Unlike the previous particular cases, increasing values of $C_r$ now lead to an increase in the imaginary part of the frequency. Negative values of the imaginary part indicate the stability of this analog black hole.

\section{Conclusions}\label{conclusions}

We have obtained analytical solutions for the radial part of the massless Klein--Gordon equation in the KABH spacetime, which are given in terms of the Heun functions. By employing the VBK approach, we got the resulting spectrum of QBS frequencies for each component of the anisotropic fluid. All QBSs are overdamped, since they have only an imaginary part. For the quintessence case, the QBSs may indicate an unstable system, since their imaginary part is positive. The same can be said for the dust matter case when the draining parameter is within the range $0 \le D < 1$, with $C_{m}=+1$. On the other hand, the QBSs may indicate a stable system for the dust matter case when the draining parameter is within the range $1 < D \le 2$, with $C_{m}=+1$, as well as when the draining parameter is within the range $0 \le D \le 2$, with $C_{m}=-1$. In addition, for the dust matter case, the QBSs may indicate a stable system when the charge parameter is within the range $-1 \le C_{m} \le +1$, with $D=2$. Finally, our solution for QBSs in the radiation case may indicate a stable system when the draining parameter is within the range $0 \le D \le 2$.

We have analyzed scalar QNMs in the context of acoustic black hole geometries by applying the WKB approximation to obtain the corresponding complex frequencies. We considered three distinct cases related to the parameter $\varpi$, corresponding to analogs of quintessence, matter, and radiation in the fluid system.
For the quintessence case ($\varpi = -2/3$), the convergence of the quasinormal frequencies was verified from the second order of the WKB approximation onward, as illustrated in Fig.~\ref{fig:FIG2_MQNs}. The analysis revealed that increasing the absolute value of the parameter $C_q$ results in a larger magnitude of the imaginary part and a higher real part of the frequency (Fig.~\ref{fig:FIG3_MQNs}). This indicates stronger damping and faster oscillations as the influence of the quintessence-like component increases.
In the dust matter case ($\varpi = 0$), the geometry presents a single horizon and the  WKB method again shows good convergence from the third order forward (Fig.~\ref{fig:FIG5_MQNs}). As in the quintessence case, increasing values of the charge parameter $C_m$ lead to more negative imaginary parts and higher real parts of the frequency, as illustrated in Fig.~\ref{fig:FIG6_MQNs}.
For the radiation case ($\varpi = 1/3$), the effective potential retains the typical structure with a well-defined maximum (Fig.~\ref{fig:FIG7_MQNs}). The WKB approximation converges even at low orders (Fig.~\ref{fig:FIG8_MQNs}). In this configuration, increasing the parameter $C_r$ results in an increase in the imaginary part of the quasinormal frequencies, as shown in Fig.~\ref{fig:FIG9_MQNs}. Although the imaginary part becomes less negative with increasing $C_r$, it remains negative throughout the considered range, indicating the stability of the perturbations.
Overall, the results show that the properties of scalar perturbations in acoustic black hole analogs are strongly influenced by the parameters governing the geometry. The quasinormal frequencies display consistent trends across different values of $\varpi$, with variations in both damping and oscillation rate depending on the chosen configuration. These findings support the utility of analog gravity models in exploring features of black hole perturbations in controlled settings.

In fact, strong-gravity tests can be conducted in the laboratory by using analog models, which are tabletop setups operating with different kinds of background; we can mention water tanks, photon fluids, Bose--Einstein condensates, electromagnetic circuits, among others. In this sense, quantum effects occurring near black holes and some classical processes remaining elusive to our (astrophysical) detectors can be, in principle, interpreted from the data of these experiments. In this work, we develop an effective geometry that describes an anisotropic fluid with draining and charge, which closely resembles the spacetime of a Kiselev black hole. The analysis of acoustic waves leads to both semi-analytic QNMs and the exact QBS spectra of this effective geometry. We hope that our results could, qualitatively or even quantitatively, align with those to be found in future experimental studies.

\begin{acknowledgments}

This study was financed in part by the Conselho Nacional de Desenvolvimento Científico e Tecnológico - Brazil (CNPq), Research Project No. 440846/2023-4 and Research Fellowship No. 201221/2024-1.
H.S.V. is partially supported by the Alexander von Humboldt-Stiftung/Foundation (Grant No. 1209836).
Funded by the Federal Ministry of Education and Research (BMBF) and the Baden-W\"{u}rttemberg Ministry of Science as part of the Excellence Strategy of the German Federal and State Governments -- Reference No. 1.-31.3.2/0086017037.
VBB was partially supported by 
Conselho Nacional de Desenvolvimento Científico e Tecnológico (CNPq) - Brazil 
- Grant No. 307211/2020
7. 
LCNS would like to thank FAPESC for financial support under grant No. 735/2024.
\end{acknowledgments}

\bibliographystyle{ieeetr}
\bibliography{ref.bib}

\end{document}